# Privacy and Twitter in Qatar: Traditional Values in the Digital World

**Norah Abokhodair**
The Information School, University of Washington
Seattle, USA
noraha@uw.edu

**Sofiane Abbar, Sarah Vieweg, Yelena Mejova**
Qatar Computing Research Institute, HBKU
Doha, Qatar
{sabbar, svieweg, ymejova}@qf.org.qa

## ABSTRACT
We explore the meaning of "privacy" from the perspective of Qatari nationals as it manifests in digital environments. Although privacy is an essential and widely respected value in many cultures, the way in which it is understood and enacted depends on context. It is especially vital to understand user behaviors regarding privacy in the digital sphere, where individuals increasingly publish personal information. Our mixed-methods analysis of 18K Twitter posts that mention "privacy" focuses on the face-to-face and digital contexts in which privacy is mentioned, and how those contexts lead to varied ideologies regarding privacy. We find that in the Arab Gulf, the need for privacy is often supported by Quranic text, advice on how to protect privacy is frequently discussed, and the use of paternalistic language by men when discussing women's privacy is common. Above all, privacy is framed as a *communal* attribute, including not only the individual, but the behavior of those around them; it even extends beyond one's lifespan. We contribute an analysis and description of these previously unexplored interpretations of privacy, which play a role in how users navigate social media.

## INTRODUCTION & MOTIVATION
The concept of privacy in the age of social media has spurred much discussion and research among scholars, designers and practitioners. While privacy in and of itself has been a topic of inquiry for centuries, the introduction and spread of digital technologies have brought novel questions to the fore that impact how privacy may be viewed and considered when we place it in the context of technology design, particularly of web-based social networking sites. When we look at existing literature on privacy, and especially in the context of technology use, we note that majority of interpretations are informed by an understanding of privacy that is based on an individualistic perspective, i.e. privacy is tethered to the notion of the self as separate from a group. To illustrate, in his book on Understanding Privacy [35], Daniel Solove provides a classification system for how to interpret privacy in six primary ways: "the right to be let alone," "limited access to the self," "secrecy," "control over personal information," "personhood," and "intimacy." While these are all bona fide and recognized approaches to understanding "privacy," interpretations that look at privacy from a non-individualistic perspective are lacking.

In particular, we recognize the gap in considering privacy as it is interpreted and enacted in the Arab world, specifically the GCC region. The GCC, or Gulf Cooperation Council, is a group of six countries that border the Arabian Gulf. Comprised of Bahrain, Kuwait, Oman, Qatar, Saudi Arabia and the United Arab Emirates, the countries share historical and cultural heritage, political interests, and economic ties. They all have a majority Muslim population who adhere to a conservative interpretation of religious texts, follow societal norms that are dictated by a patriarchal system, and follow Islamic—or Sharia—law. Importantly for this research, the concept of privacy is held within high regard among citizens of the GCC. Privacy is a prominent value that touches upon all aspects of behavior; from how people dress, to the design of homes, to their interactions in physical and digital environments.

We focus on the discussion of privacy and related terminology within digital environments by Qataris. In particular, we look at discourse posted on Twitter, a popular and widely-used social media platform in Qatar. Our aim is to understand how a public, wide-reaching social media platform is used to discuss ideas and circulate recommendations regarding "privacy" within a culture that views privacy as a sacrosanct right. Due to the similarities between Qatar and the other GCC countries, our findings extend beyond Qatar to include the wider region.

We examine tweets posted by Qatari nationals who contribute to the conversation on privacy from a local, or regional, perspective. We focus on Twitter because it provides a lens through which we are able to observe how privacy is mentioned and discussed, and how Twitter users come to define and revamp what privacy is via Twitter. Qatari society is rapidly evolving; there is great pressure on the government and the population as a whole to modernize, while at the same time, to preserve tradition. Observing and understanding how privacy is interpreted via social media provides us with the ideal backdrop through which to understand this tense movement. By tapping into the rich corpora comprised of thousands of tweets, we explore the meaning of privacy using a mixed-method approach that involves quantitative and qualitative techniques.

### Non-Individual understandings of privacy

According to Solove, "privacy involves one's relationship to society; in a world without others, claiming that one has privacy does not make much sense." Taking a cue from [28,35], we provide a contextually grounded understanding of privacy based on discourse that takes place via Twitter. In this communicatory milieu, notions of privacy are formed through a negotiation between individuals and society. This intersubjective formation of privacy falls in line with Altman's Privacy Regulation Theory, in which he argues against the meaning of privacy as total withdrawal, and instead advocates for understanding it as a process of optimization [6,7]. In other words, privacy is attained as individuals arrive at the acceptable personal balance between withdrawal and disclosure to a group. In performing this study, our aim is to introduce an understanding of privacy that has yet to be theorized to any great extent – i.e. by citizens of the GCC.

## BACKGROUND

### Understanding privacy in Islam

In the Arab world in general, Islam is the main force that dictates the institutional norms, patterns, and structures of society. The primary sources for the legal framework are the Quran and the 'Hadith' (verbatim quotes of the Prophet Muhammad). Together, they establish the foundation of *sharia law*, which is widely practiced throughout the region. When Muslims speak of "privacy," they are often referring to a host of values that are foundational to both religious and cultural norms. Islam recognizes the importance of the fundamental human right to privacy [19]. In the Quran, the mention of privacy arises in a few passages—one that prohibits people from spying and gossiping on one another: 'Do not spy on one another' (49:12), and the second mention in the commandment to seek permission when entering another's home: 'Do not enter any houses except your own homes unless you are sure of their occupants' consent' (24:27).

For Muslims, there are many domains in which privacy should be sought or given. In reviewing the literature, we note that there are three domains that are often mentioned vis-à-vis Islamic practice and privacy, these are: (1) privacy of the home, (2) privacy for gender exclusive spaces and gatherings, and (3) individual privacy.

#### *Privacy of the Home*

When dealing with privacy in the home, multiple layers are implied. At one end is the privacy of the entire property; a need to protect the home from outsiders. At the other end is the protection of privacy amongst the members of the home and guests [19, 29, 33]. Figure 1 illustrates these, in addition to the varying degrees within. The reader will see that females' privacy is of profound importance and as a result they are afforded more privacy than males. Then comes the privacy of the family which is an important aspect of Islamic teachings. At the heart of the hierarchy we find privacy of the self.

It is important to note that the concept of a nuclear family is not commonly used within Muslim communities as the care for the family members includes the extended family as well [21]. This is one aspect of collectivistic and relationship-based societies that is of paramount influence on how people conceptualize privacy.

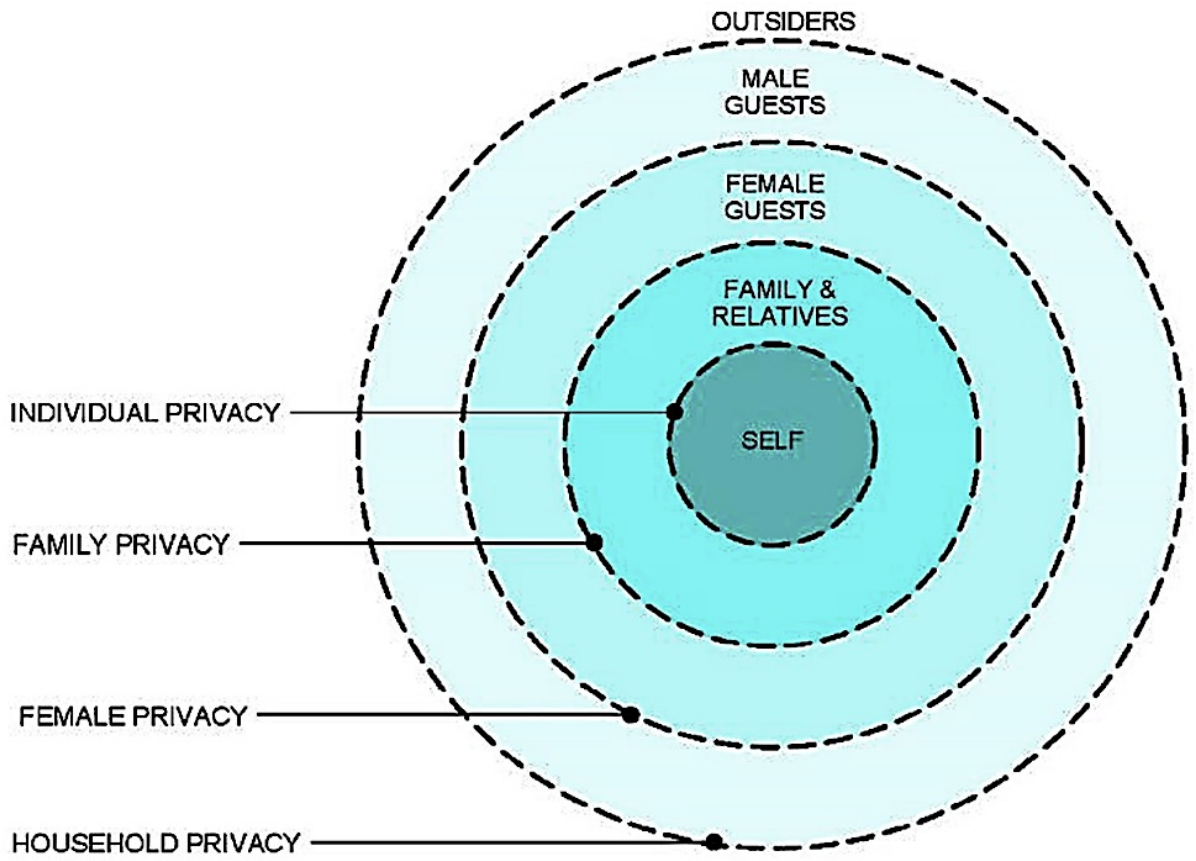

**Figure 1: Hierarchy of home privacy domains [8]**

Researchers who focus on privacy in Arab Gulf homes note how religious and cultural values manifest in architectural design. In their study, Sobh and Belk provide detail about traditional home design in the Gulf Arab countries, where houses are typically designed with an inward-facing center to protect the family from the public eye [33]. The idea is to maintain the sanctity of the home, which is considered sacred and pure, and which must be guarded from the gaze or intrusion of non-family members. In another study, Othman et al. discuss the issue of privacy and design in Muslim homes by conducting an extensive literature review [29]. They found that privacy in Muslim homes can be attained through the maintenance of three important types of privacy: visual, acoustical and olfactory privacies.

#### *Privacy for Gender Exclusive Spaces*

Privacy among single-gender gatherings inside the home as well as elsewhere is also of paramount importance. In the GCC, there is great emphasis on the need to respect the privacy of male or female-only spaces. There are two kinds of relationships that guide the interaction amongst opposite genders within the family and outside: *mahrams*, this group consists of unmarriageable kin (e.g., father, son, brother) and *non-mahrams*, men and women who can marry each other. There are clear rules that guide the interactions and socialization between opposite genders. For instance, women are not required to wear *hijab* (a veil that covers the head and chest) in the presence of mahram men.

### *Individual Privacy*

The preservation of one's chastity is of a great importance in Islam. Multiple verses in the Quran and Hadith describe the importance of individual privacy that is mainly concerned with the protection from the intrusion of others, particularly the protection of the intimate parts of one's body. This practice is represented by the word *awra* (عورة) which are the parts of the body that must be covered by clothing[1]. In Islam, the protection and respect of one's awra is a personal responsibility, as well as that of society as a whole. For example, one must lower their gaze when another's awra is exposed. The mention of awra in the Quran is associated with the rule to seek permission before entering a private space so as to avoid walking jn on someone, and exposing their awra.

The protection of awra is often associated with *hurma* (حرمة). Hurma literally means "holiness," or "sanctity." It can also be used in reference to a woman, a sacred space (mosque or home) or a sacred time (holy month). Essentially, hurma is the law that protects anything that is unlawful to obtain or look at without permission. Risking intrusion on the sanctity (hurma) of places and people is considered sinful [29,33]. Consequently, respecting others' privacy, and maintaining one's own privacy is a rule that citizens of the GCC adhere to by seeking blessings of God and maintaining a respectful image that is largely accepted by society.

Thus far, we have stressed how the protection of privacy is tantamount to following the example of the prophet Mohammed; what this process looks like in the age of social media is cause for new interpretations and implementations regarding privacy.

## Study Context

We focus on the Twitter communications among Qataris and those living in Qatar to capture how privacy is discussed, enacted and taught in this part of the world.

### *Population*

Qatar is a small peninsula that juts into the Arabian Gulf. It is home to a native population of approximately 300,000 Qataris, who live among a total population of 2.3 million [32]. With the majority of the population comprised of expatriates, national identity and observance of cultural norms hold great importance among Qataris.

[1] The instruction on which body parts must be covered varies between different schools of Islamic thought.

### *Technology Adoption*

Internet penetration is high in Qatar; 85% of the population has Internet access. In addition, social media are popular and widely used among the native population—65% of Qataris have an Instagram account, 44% are on Facebook, and 46% use Twitter [30].

### *Gender*

Qataris use social media platforms frequently, but among this population, there is a supposed gender imbalance. The latest figures show that 28% of Qatari Facebook users are female [12]. However, it is entirely plausible that this figure does not accurately account for users' gender. Based on our interactions with participants in previous research, [1] we know that many women prefer not to identify their gender when they create social media accounts. Due to the stigma and potential for bringing shame onto one's family associated with female visibility on public platforms, it is often the case that females either leave the "gender" field blank, or say they are male [18]. Therefore, knowing the gender breakdown of Qatari social media users is a difficult, if not impossible figure to determine automatically.

### *Legal Approach to Privacy*

The ways in which the Qatari legal system treats privacy is additionally helpful in explaining how privacy is approached and understood by the native population. The law that protects an individual's privacy stems from religious and cultural expectations; it is not a civil right. Said law dictates how personal matters ought to be treated and respected. The Penal Code sets out certain rights of privacy (haq fe al-khososyah) that include:

"The sanctity of the individual's privacy shall be inviolable, and therefore interference in a person's privacy, family affairs, home or correspondence, or any other act of interference that may demean or defame a person, shall not be allowed, save as permitted by the provisions stipulated in the Law" [4].

Additional articles within Qatari law mention privacy, though it is never explicitly defined—it is treated as something that is intertwined with multiple facets of one's life. In a Qatari context, privacy is bound up in the most fundamental and revered details that comprise human existence as seen from an Islamic perspective. It is a sacrosanct value that must be respected and upheld.

When it comes to social media use, Qataris turn to legal articles that speak to the use of digital technology vis-à-vis privacy, though not in so many words. Laws that mention privacy in relation to news or pictures—such as Article 331—state the penalties for "spreading news, photographs, or comments related to secrets of the private life of families or individuals, even if they are true." i.e. regardless of veracity, if you are to share private information about another person or family, you will be punished. In addition, Article 333 lists the following acts which are considered illegal intrusions into others' private lives if carried out without their consent: opening a private letter or telegraph addressed to someone else; listening secretly to a telephone call; recording or transmitting conversations that took place in a private place through any kind of apparatus; and taking or transmitting photographs of an individual in a public place through any kind of apparatus.

It is conceivable that violating privacy through actions taken via social media could be punishable by Qatari law, and perhaps more seriously, could severely harm honor and bring shame to an individual and their family.

## RELATED WORK

### The Role of Social Media in the Arab World

The use of technology in Arab World has been investigated during the recent Egyptian, Tunisian, Yemen, and Libyan revolutions, and the ongoing Syrian Civil War (the so-called Arab Spring) [2,3,9,23,26]. Researchers especially focus on the innovative ways Arab youth utilized Social Networking Sites (SNS) during that time [23]. They highlight the opportunities offered by digital media for youth to participate in collective action, such as creating Facebook events for demonstrations, and live tweeting/blogging public action. Others focus on the critical roles of digital societies—groups of stakeholders collaborating via electronic means—in the organizing, coordinating, and publicizing of the protests [9]. Different platforms have been utilized as a medium to express opinion and to call for equal rights [3]. As noted by [23], Facebook enhanced the ability of activists and protesters to coordinate peaceful protests while allowing larger segments of the public to participate as citizen journalists documenting and sharing witness accounts and cell phone images and videos. Another study by [9] analyzes the different roles of Facebook during the protests that led to the resignation of President Mubarak in Egypt and the start of a new sociopolitical period. While the results of these studies show the crucial role these technologies had on Arab youths' political awareness and engagement, they also revealed a complex set of practices and tensions, especially in how they experience privacy and manage identity.

In this research, very little has focused on studying the daily interactions and experiences with digital technologies, and the opportunities and challenges these technologies bring in terms of managing privacy, honor, and identity.

### Cross-Cultural Privacy

Privacy is multidimensional and contextual; it is interwoven with notions of identity, seclusion and autonomy. According to [14,27], age, gender, education, and culture are the most important factors that affect online privacy concerns among individuals. A survey study conducted by [24] reinforces cultural norms as a major factor. Participants in India reported fewer concerns with providing personal health information online and lower levels of privacy concerns compared with American participants. In the Arab world, a questionnaire of 325 social media users in the Emirates and Egypt [27] revealed no personal restrictions on sharing personal information, such as city, religion and gender. However, respondents do not share family photos, cell phone numbers, and friends' photos. In addition, [27] reports that the more respondents were concerned with their online privacy, the less likely they were to give accurate personal information. These results have been discussed in the privacy literature as an example of *data withholding* (i.e., limited disclosure or removal of data) [35] or *data fabrication* (i.e., misrepresentation of data by providing inaccurate or incomplete information) [25].

The difference in gender roles and expectations within a society is another key aspect to understanding privacy. A study by [1] provides examples of the different ways women from Qatar and Saudi Arabia protect the privacy of their social media accounts based on adhering to societal expectations. For example, women tend to not use their real pictures when creating social media accounts because this might create negative consequences regarding their own and their family's reputation. In addition, the profile photo might risk exposing a women's awra (in this case, their face) to the public.

Much research has been done within the field of Human Computer Interaction (HCI) that provides insight for technology designers and policy makers to promote privacy. However, the notions of privacy that are inscribed in most technologies and services derive from a particular perspective—privileged, technologically-oriented, and North American/Western European [16,22]. One way to address the lack of diversity regarding privacy in technology design is through the inclusion of a wider spectrum of users, and the employment of methods that explore differences across cultures by keeping myriad values and beliefs in mind during the design process. Many researchers have contributed work in this area [5,14,15,30,37].

Challenging the Western norm, a recent content-driven study by [38] provides new insight into how privacy is conceptualized in China. The authors conducted a semantic network analysis of 18,000 posts on Sina Weibo, China's largest social networking platform. They focused on posts containing the equivalent word to 'privacy' in Mandarin. The individuality that was expressed in the study was different than the one commonly referred to in cross-cultural research. In China, the individual is expected to "still engage in constant dialogue with the traditional Chinese self embedded in a network of relations and situations."

Our study expands on the results of [38] using automated and manual analysis of Twitter discourse, as we aim to explore the meaning of privacy through discourse on Twitter among Qatari users.

## DATASET CREATION & REFINEMENT

The analysis below is conducted by the authors, who have all been living and conducting research in the GCC for several years. In addition, the team includes two native Arabic speakers. We aim to answer the following research questions based on three types of analyses:

*RQ1) How do Qataris express/refer to privacy in Twitter communications?*

*RQ2) What topics or contexts do they refer to in their discussions/discourse?*

*RQ3) Are there any indications of gender differences? Is gender a topic that is discussed regarding privacy?*

### Step One: Identifying the Qatari Users

We began with a dictionary of locations in Qatar including the main cities and districts. We then matched this list to a 45 day sample of the Twitter Decahose (a sample of 10% of all Twitter traffic.) For each user whose tweet was captured, we requested the list of their followers (based on the assumption that in this region, followers are more likely to be from the same country). Then, we requested the profiles of the followers and filtered out those users who did not mention a location corresponding to Qatar. The same process was iteratively executed on the list of newly added users; i.e. find their follower IDs, then follower profiles, then filter out followers with irrelevant locations, until no new users are identified.

The resulting list consists of over 100K users who claim to be living in Qatar. In order to identify the Qatari users, we first built a dictionary of the most common family names in Qatar, both in Arabic and English with different spellings[2] (such as Al Hammadi, Al-Hammadi, as well as the Arabic original الحمادي). Using this list, we identified the group of users whose name or screen name matches a Qatari family name. The result of this operation is a list of 12K Twitter accounts. Finally, we gathered the latest 3200 tweets of each of these users (the maximum allowed by Twitter API). This collection generated a dataset of over 8.5M tweets that were posted between September 2006 and July 2014.

### Step Two: Identifying Privacy Related Keywords

Following [28,35], we aimed to capture related dimensions of privacy as it is conceptualized and practiced in face-to-face interactions, in addition to privacy in the digital world. Building on the literature and our understanding of the Arabic language and Qatari culture, we focused on mentions of privacy that are related to honor, reputation and individual sanctity. The generated list of Arabic words related to privacy is explained in Table 1.

| Arabic word | Word in English alphabet | English equivalent | Number of Tweets |
|---|---|---|---|
| خصوصية | Khsosyah | Privacy | 388 |
| عزلة | Ozla | Isolation | 448 |
| سرية | Sereyah | Secrecy | 839 |
| حرمة | Hurma | Sanctity | 424 |
| عرض | ‘Ird | Honor | 9660 |
| شرف | Sharaf | The space where honor applies | 5757 |
| حياة خاصة | Hayat Khasah | Private life | 3 |

**Table 1: List of words related to privacy**

[2] The list of names is available at https://tinyurl.com/qatarinames

In Arabic, there is no single word that means “privacy”— it always depends on the context in which one is trying to convey a sense of privacy. The most general Arabic word that indicates privacy is *khososyah* (خصوصية). If we translate “privacy” from English to Arabic in Google Translate the results will include isolation (عزلة), and secrecy (سرية), as well as khososyah (خصوصية).

In generating Table 1, we took into account eight parts of speech (e.g. nouns, pronouns, verbs) that expanded the list to include all subsets of each entry. Our list contained 26 words and terms: خصوصيتي, خصوصية الخصوصية, عزلة, العزلة, عزلتي, سرية, السرية, سريتي, حرمة, الحرمة, حرمتي, حياة خاصة, الحياة الخاصة, حياتي الخاصة, شرف, الشرف, شرفي, اختلاط, الاختلاط, إختلاط, الإختلاط, عرض, العرض, عرضي.

A sub-string matching algorithm that matches each of these 26 keywords to tweet text resulted in a collection of 18,233 tweets. In Table 1, we provide the size (in number of tweets) of each category (aggregated over parts of speech).

### Step Three: Identifying Relevant Tweets

Next, we turned to the tweets in which these keywords appear. We use frequency analysis and manual inspection for smaller collections, and algorithmically-assisted sub-topical analysis for larger ones.

#### *Manual Frequency Analysis*

We assessed tweet content by examining the top 100 words (after removing Arabic stop words) for each group, and then performing a high-level manual analysis. As a result, eliminated the categories *sereyah, sharaf,* and *ozla* because these tweets were mostly not related to the concept of privacy, or were ambiguous. Other collections, such as those about *hurma*, *hayat khasah* and *khososyah* were merged, as we noticed topical similarities.

#### *Latent Dirichlet Allocation (LDA)*

For the remaining term which resulted in a collection of over 9K tweets—*‘Ird*—we used an automatic topic discovery algorithm to disambiguate its several meanings, and selected only topics relevant to privacy. We did this to classify tweets into different semantic categories, and hence identify the relevant subset of tweets. To infer topical categories, we employed the Latent Dirichlet Allocation (LDA) [10] algorithm, which attempts to explain similarities among groups of observations within a dataset.

The result of LDA was a set of 50 topics, each of which was identified as a list of words extracted from the tweets, and a probability matrix stating the contribution of each topic to each document. For the sake of simplicity, we assign only one topic (with the highest probability) to each tweet. The resulting collection included just over 1772 tweets, which were then further analyzed.

## DATA ANALYSIS

Taking an empirically-based, ground-up approach, we analyzed 2587 privacy-related tweets, identifying major themes relating to privacy concerns that involved both digital and face-to-face situations. We employed both quantitative and qualitative approaches throughout the analysis; the two Arabic speakers regularly discussed the analysis, compared notes, and came to mutual understanding regarding results and claims.

### Quantitative Analysis

We begin with computing co-occurrence between the terms in the context of ‘ird and khososyah. We then generate graphs with terms as nodes and extent of co-occurrence as weight of edges, as shown in Figures 2 and 3. The size of the node’s label reflects the frequency of the corresponding term in the dataset, and the width of the edges provides an approximation of the number of times the two terms appear together.

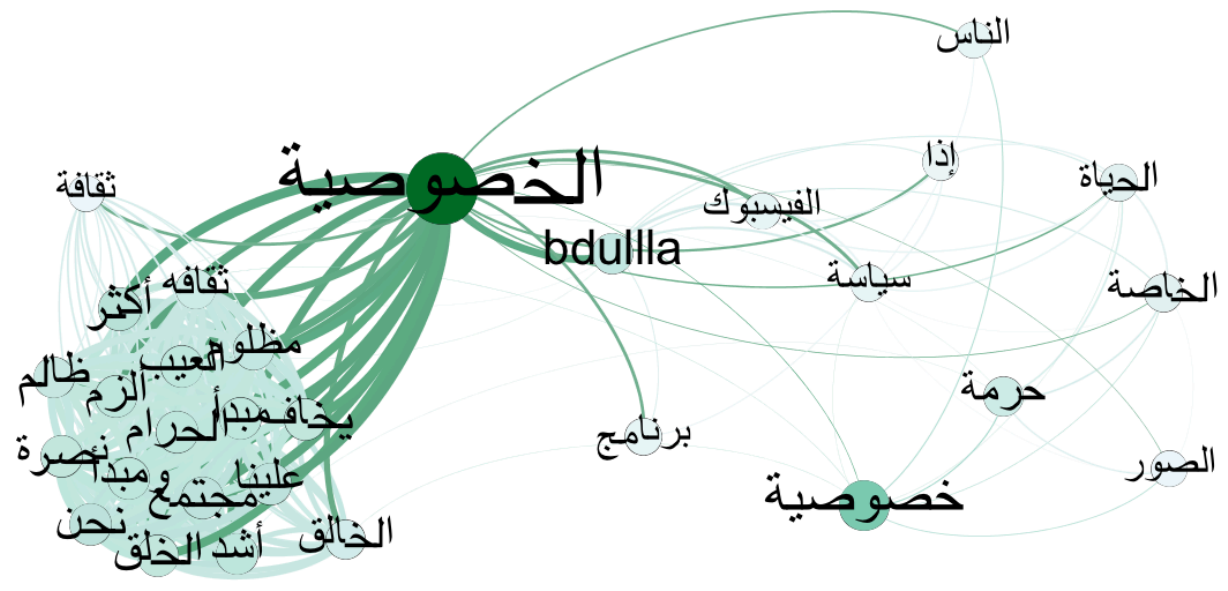

**Figure 2: Terms Associated with Khososyah (privacy)**

There are two ways in which the term khososyah is mentioned in the dataset, both with and without an article preceding it. In Figure 2, we observe that the central node—Al-Khososyah (“*the* privacy”)—is mainly associated with terms such as *fear, culture, demerit, society, haram, software, Facebook, policy,* and *people*. While the second most frequent term, khososyah (privacy *of*) is associated with terms such as: *photo, hurma, people*, and *life.* While the first set of terms is concerned with defining the concept of privacy, the second set is primarily concerned with the context, or the space, to which privacy should apply.

In Figure 3, the central and most frequent term is ‘ird, which is connected to three clusters. ‘Ird—or honor—is of the upmost importance regarding the protection of privacy [11]. The top cluster mentions topics/descriptions such as: *picture, personality, suitable, values* and *ambassador.* The tweets associated with this cluster talk about the photo as a representative of the user, instructing users to be mindful of the image they put forth. The left cluster revolves around awra, a term which translates as “private parts of one’s body,” but which carries deeper meaning, and extends to the notion of anything that one should or wants to keep private.

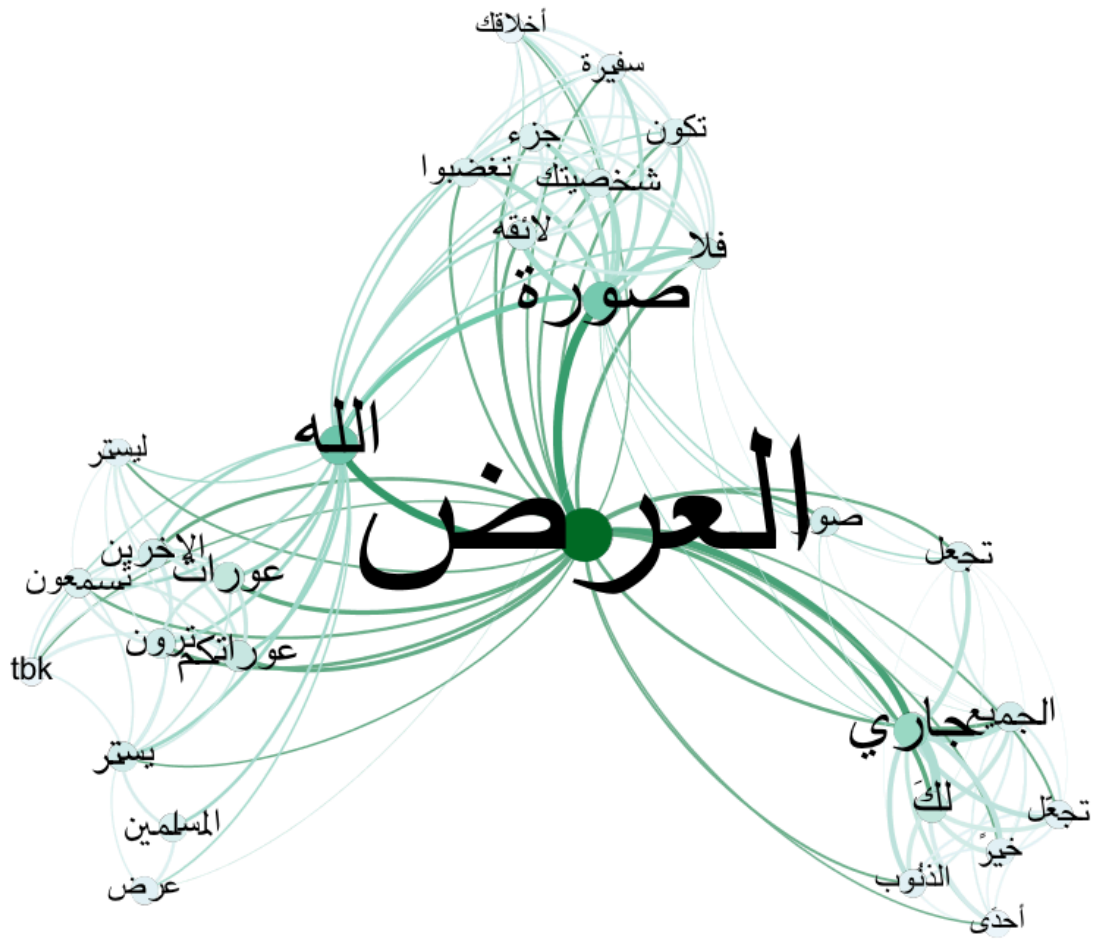

**Figure 3: Terms associated with ‘Ird (Honor)**

Tweets that mention awra discuss anything that should not be disclosed in public, be it parts of the body, certain behaviors, or thoughts. The terms in this cluster are *awra, others, your awra, Muslims,* and *protect/hide*. Most of these terms are linked to the term *God*, indicating a religious connotation. Overall, the main idea expressed in these tweets is a call to protect others' awra if one wants God to protect their own awra, i.e. “do not violate others' privacy if you do not want your own privacy to be violated.” In the lower right cluster, we see words that relate to the idea of accumulating good deeds. In Islam, there is an understanding that deeds, be they good or bad, follow a person and their family through eternity; i.e. the death of a person does not mean death of their actions (an important point we discuss later).

### Discourse Analysis of Privacy Related Tweets

To better understand what and how Qataris communicate when discussing privacy, the Arabic speaking authors read each tweet in the dataset, with the goal of identifying themes—a process which utilizes computer-mediated discourse analysis (CMDA). CMDA is defined by Herring as: “any analysis of online behavior that is grounded in empirical, textual observations.” In performing CMDA, analysts take the technological affordances of computer mediated communication systems into account, and adhere to three assumptions as listed by [20]:

1. Discourse exhibits recurrent patterns
2. Discourse involves speaker choices
3. Computer-mediated discourse may be, but is not inevitably, shaped by the technological features of computer-mediated communication systems.

By taking a CMDA approach, we aim to answer macro-level questions that take an entire society or culture into account, but which can be explored by performing detailed, micro-level analysis.

### Qualitative Coding

We were careful to take context into consideration when interpreting and formulating themes, and eventually identified a set of topic categories of increasing specificity that speak to the concepts or ideas that are mentioned in the data. This process involved several iterations of coding which we describe below.

#### *First Round: Coding For Relevance*

In the first round of coding, the Arabic speakers read all tweets in each file (‘ird = 1772; khososyah = 815) and coded for *relevance*, meaning that we decided whether or not the tweet was privacy-related. We found that in some cases, tweets used a similar word to *khososyah,* but in reference to unrelated concepts. For example, “khososyah” is used to describe “private teaching lessons;” a topic outside our area of inquiry. Another source of ambiguity is the short vowels (*harakat*), which in Arabic can change the word meaning, and which many platforms, including Twitter, do not support. For example, the word ‘ird could be associated with the word “day,” meaning “judgment day,” which is not directly relevant to our research. Coding for relevance brought the set of tweets from 2587 to 354.

#### *Second Round: Development of Primary Codes*

Next, we use CMDA, developing of a list of qualitative codes that provide a meta-level description of the tweets. The purpose of the second pass was to determine what the relevant tweets are communicating at a more detailed level. The Arabic speakers developed the list of codes through multiple rounds of reading and discussion amongst the team. The codes used in this phase are not mutually exclusive; annotators may assign more than one code to a tweet. The primary codes are as follows:

*Individual:* Content refers to personal privacy

*Society:* Discussion of society as a whole vis-à-vis privacy

*Family:* Reference to family members, tribes

*Technology:* Technology-related mentions of privacy, i.e. technology companies, cellular carriers, and/or applications

*Law:* Information about the legal aspects of, or rights to, privacy

*News:* Communication about media coverage of privacy and technology use and their impact in the country/region

*Religion:* Focus on references to God, the afterlife, or sin

*Culture:* Cultural aspects and references, in contrast to religion, e.g., reference to non-religious proverbs

*Personal Identifiable Information (PII):* When the subject matter is related to some kind of personal data that is stored, transferred or otherwise used via technological means

*Opinion:* Personal views about privacy. Including (but not limited to): opinions; jokes; poetry; praise for privacy as a societal need; criticism of the modernity that is changing views of privacy; rhetorical questions

*Warning:* Warnings or cautionary statements regarding the maintenance of personal and others’ privacy

*Recommendation:* Instructions and suggestions on how to protect privacy for oneself and others

*Concern:* Expression of worry or concern over privacy violations or lax attention toward privacy

#### *Third Round: Specification of Topics*

As annotators were coding tweets with one or more of the above categories, they noted that further specification was helpful, i.e. labeling tweets with the codes listed above was useful, but there were more details within the tweets that would benefit from another tier of analysis. Throughout the second round of coding a second tier of codes was developed—we frame these codes as “topic codes,” i.e. they are more detailed, micro-level descriptions of content. These include applications like Facebook, Twitter, WhatsApp, web browsers such as Chrome or FireFox, and other higher-level concepts such as respect, reputation, morality, hurma, God. In addition, personal and local topics and concerns were mentioned, such as Qatar University, gender, profile photos, and surveillance.

## DATA DESCRIPTION

Below, we describe the results of our qualitative analysis. In Figure 4, we illustrate the distribution of the most frequent qualitative codes. Nodes in red are *primary codes* and nodes in grey are *topic codes.*

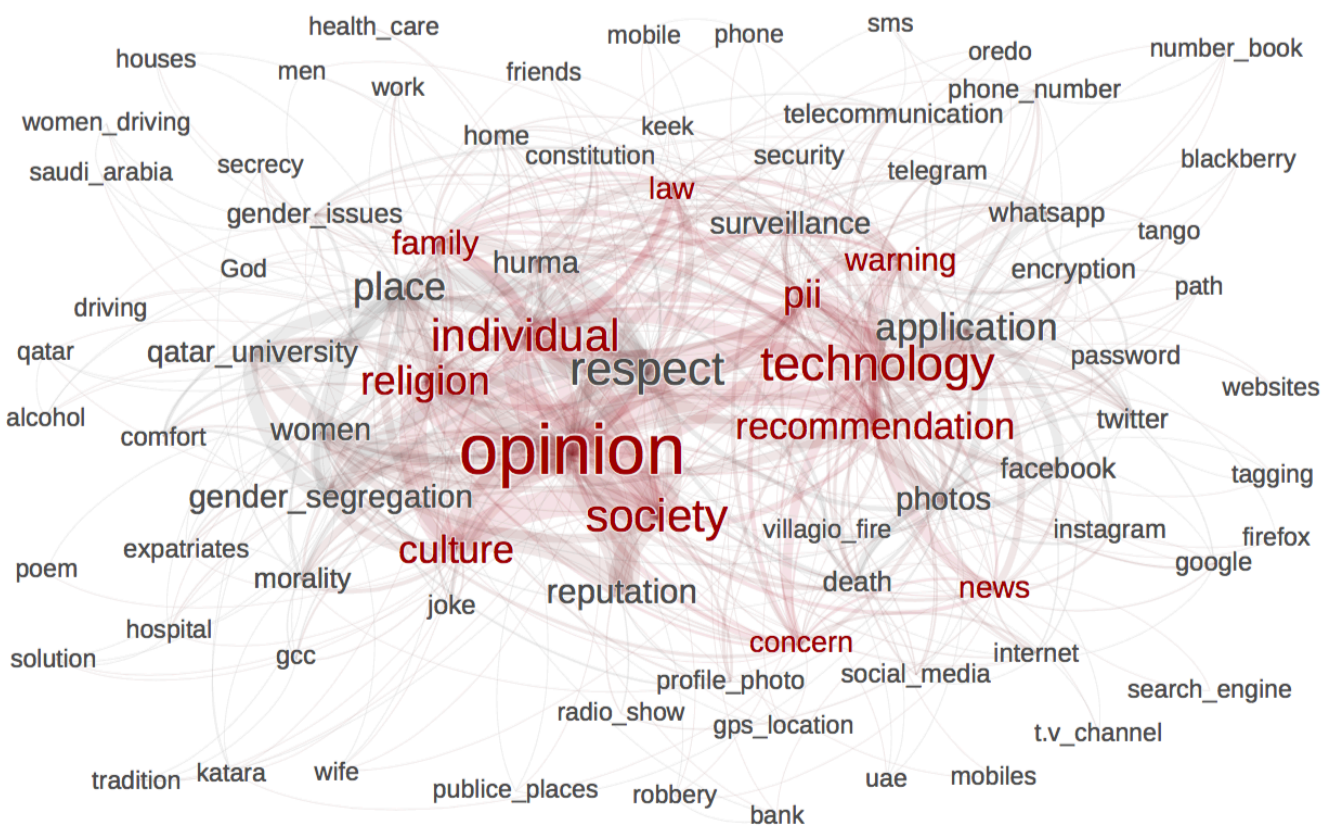

**Figure 4: Code co-occurrence network, primary codes in red**

### Concerns about Technology Use

Twitter users were overwhelmingly concerned with how to safely and effectively use new technologies while maintaining a sense of privacy and adhering to societal expectations. Along these lines, many users in our dataset provide warnings and recommendations about applications and other digital technologies that could jeopardize privacy. They also express worry about the growing adoption of social media applications. Technological references mentioned include social media applications, web browsers, telecommunication companies, the government ministry of ICT, and encryption.

### *Meta-level Privacy*

One way in which users discussed their misgivings about maintaining privacy while using technology was to express concerns regarding the mishandling of personal identifiable information (PII) through poorly managed databases and social networks. To illustrate, one user posted the following tweet:

*What [company name] is doing towards their customers by selling their PII needs to stop and we have to stand against this company that doesn't respect their client's privacy.*[3]

As a response, the company followed up with multiple tweets explaining the situation:

*This information is not true. The telecommunication law does protect client privacy.*

*The telecommunications law stands for the preservation of customers' privacy and we can give this information only to legal authorities.*

In this case, Twitter was used as a platform to bring the technology service company and the clients together to clarify issues regarding information privacy. While taking to Twitter to carp about a service provider is hardly unique, we highlight this example to stress the importance of the company's timely response, succinct and careful articulation regarding the security of client data, and citation of legal code. Because privacy is of paramount importance, the company is obligated to not only reassure clients, but to support their position in a way that invokes not only law, but a formal tone that indicates the company understands the gravity of the accusation.

In addition, we noted that it is common for users to express fear when technological trends are first introduced, such as new social media applications. They are unsure what negative consequences may ensue if they or their family members use them. However, ironically, we noted that Twitter was a platform users employed to convey Twitter-related privacy issues.

### *Privacy Invasion Warnings*

Adherence to privacy in the GCC stems from a need to adhere to religious and cultural norms. We note that people often reference punitive measures and/or God's wrath when discussing the infringement of others' privacy, whether in digital settings (e.g., stalking on social media) or face-to-face (e.g., spying on another's home). For example:

*Who reads a message/personal note that is meant to be for someone else without their consent then they have assaulted the sanctity (hurma) of that individual's life and can get up to one year in prison as per the law. So beware...*

In this example, the reference to personal privacy is through the use of the term *hurma*, which is referring to the Quranic rule that requires one to seek permission before entering another's personal space. In this tweet, the permission rule was invoked in terms of accessing another's email..

In regards to warnings of privacy in public spaces such as restaurants, people used Twitter to spread warnings of how marital privacy may be violated:

*Do you think your privacy is protected when you are with your wife in the restaurant family section partitioned with frosted glass barriers? http://t.co/q7pUmKo6ll*

This tweet is referencing gender segregation, which is often practiced as an unspoken law in some public places in Qatar. Often, restaurants will separate an area from the larger dining room with frosted glass, which allows women who cover their face with a *niqab* (a cloth that covers the face) the option to remove it so they can freely eat. This Twitter user is warning people about a simple trick (shown in the link) that involves applying scotch tape to frosted glass, which makes it transparent. If used, this trick risks the exposure of *awra*, in this case, a woman's face [31].

## Privacy and Surveillance

Our data revealed interesting results regarding the discussion of surveillance. People not only discussed government-led surveillance, but also highlighted surveillance performed by peers.

*People, we appreciate your concern during the current events, however, families still have their privacy and hurma that need to be respected and it is better to stay away from people's business.*

In another tweet, a person is sharing his opinion regarding the phenomenon of *social surveillance,* which is commonly practiced in relationship-based cultures, of which Qatar is one [37]:

*We should not restrict others' right to freedom of expression and we need to have the laws to protect this right, but freedom is a matter that is only granted when others' privacy is respected.*

In this example, the tweet author is engaging with two values: *privacy,* and *freedom of expression,* to explain that they are closely related. The author is suggesting that to achieve the goal of freedom, members of society need to respect the privacy of each other and not infringe on others' personal matters. In the Gulf, the notion of privacy—as revealed in our data—is negotiated amongst the group; it is not something that individuals are able to seek out and attain without societal consensus. Both those who watch and those who are being watched are empowered to arrive at a mutual understanding regarding how social media should be used vis-à-vis the maintenance of privacy.

[3] All tweets are translated by the first author from the original Arabic into English.

### Privacy and Honor: Immortal Values

Another issue that arose was that of photos of dead people. The mentions often refer to the sanctity or hurma of the dead person. it is a sin to impinge upon their privacy, even though they are dead:

*Violates your privacy in your grave: Facebook updates its privacy policy to allow your friends and family to access your profile information and pictures after dying upon their request.*

This tweet brings to light an aspect of Islam that is linked to eternal life. For Muslims, values such as privacy and honor have an everlasting association, so whether alive or dead, privacy and the protection of it remains the individual's responsibility. This concern is also seen in discussions regarding the sharing of dead people's photos. In the following example, the issue of concern is the spread of photos of the dead after a fire that resulted in the death of nineteen people.

*Death has its sanctity, please respect it by refraining from sending photos of dead people. I see that some of them began to circulate on Twitter. #Fire[mall name].*

#### *Profile Photos*

When conducting qualitative coding on the 'ird file, we noted that the most discussed subject was profile photos. In this case, the words *'ard* (display) and *'ird* (honor) are both spelled in Arabic in the same way with different harakat, which changes the meaning of the word. This resulted in the appearance of the "display photo" discussion in our dataset, although unintentionally. This was an unexpected opportunity, as it expands our understanding of how people conceptualize their privacy and honor regarding digital media. The display photo was clearly a source of discomfort for many, especially for women wanting to use their real photos. The following example illustrates how people viewed the profile photo and how it was closely related to morality and faith:

*Your display photo is part of your personality and is the ambassador of your morals. Therefore, do not upset God with an inappropriate profile picture.*

The message here is that profile pictures are part of one's personality (i.e. it plays the role of an ambassador as it represents who one is), thus, one should select a suitable photo, so as not to provoke God's wrath. The following is another example of the significance of profile photos:

*Display pictures of women is one of the ongoing sins. So if you do not create yourself an ongoing source of good deeds then do not create the opposite.*

This discourse around profile photos illustrates how privacy and honor are intertwined; they are communal values that are negotiated and agreed upon by all members of society. Again, this is significant because privacy and honor are considered immortal values that do not end or stop existing when one's life ends.

#### *"Exposed to" Versus "Exposing"*

Qataris often took to Twitter when situations of discomfort happen. A sensitive topic that appeared in our dataset concerned the adherence to dress codes. This led to a country-wide campaign to educate tourists and foreign residents to respect the country's dress code [13].

*On the lack of modesty amongst expatriate women and men, and the lack of respect of our Arab-Islamic society privacy [link].*

Reactions to this campaign were divided between support and discontent. Supporters often cite the saying "When in Rome do as the Romans do" and on the opponent side people expressed discomfort with the lack of tolerance.

In this example, the use of the term *khososyah* is associated with privacy. However, it draws upon a different facet of privacy—one that is not completely in the hands of the individual. In the GCC, privacy is not merely what one exposes of themselves, but also what they are exposed to (what other expose). So, if a Qatari is exposed to another's uncovered body, this is seen as an invasion of privacy. Again, privacy in this sense is negotiated between society as a collective and individuals, and the expectation is that a common enactment of privacy will materialize. We theorize that when privacy is conceptualized as a collective value it impacts the way people perceive and enact it. So, when an individual is thinking of their own privacy and how to protect it, they are thinking of others' privacy as well.

### Gendered Expressions of Privacy

To answer the question, "is privacy gendered?" we conducted a third level of analysis by looking at users' gender.

#### *Gender Labeling*

Gender identification of social media users has attracted a lot of attention in recent years [17]. While most approaches identify gender by looking at content and social ties, the most practical ones try to identify gender directly from users' names and screen names. For instance, one widely used tool is the Genderize[4] API, which performs gender classification based on user names. The API uses a database of names extracted from major social networks, and produces the most likely gender associated with a first name, which can be *male, female,* or *none* when the gender cannot be detected. However, given the nature of names people provide in online platforms, it is common to end up with a high fraction of unknown gender (none). In [34], authors reported that over 30.7% of the US based Twitter accounts they used were labeled as "none" using the Genderize API. Given that we deal with Qatari users, identifying gender is even more complex for the following reasons: (i) user names in the Arab world are less likely to be correctly identified as they are often spelled as a combination of Arabic and Latin characters (Arabeasy), (ii) Women tend to anonymize their names and gender in their online personas as means to protect their privacy [1,5,18]. Thus, we decided to replace the automated gender labeling process by manually labeling the users in our dataset.

---

[4] https://genderize.io/

The manual labeling was done by the same authors that conducted the manual annotation. They looked at each user's name, screen name, and link to their Twitter profile. While in some cases, names and screen names were enough to assign gender, most accounts required more in-depth analysis into their profiles (profile picture, tweets, and biography.) Annotators relied on their cultural knowledge to infer the correct gender. Given the study context and our knowledge of previous studies that report statistics with low percentages of women [12,30], we had a hypothesis that men would dominate in our dataset. Surprisingly, we found that a substantial fraction of users who took part in the privacy discourse are females (47%). On average, women contribute 1.367 tweets while men contribute 1.445 tweets. We also identified two accounts corresponding to families (tribes) not to individuals; these accounts were labeled as groups and discarded from the analysis.

### *Gendered Privacy*

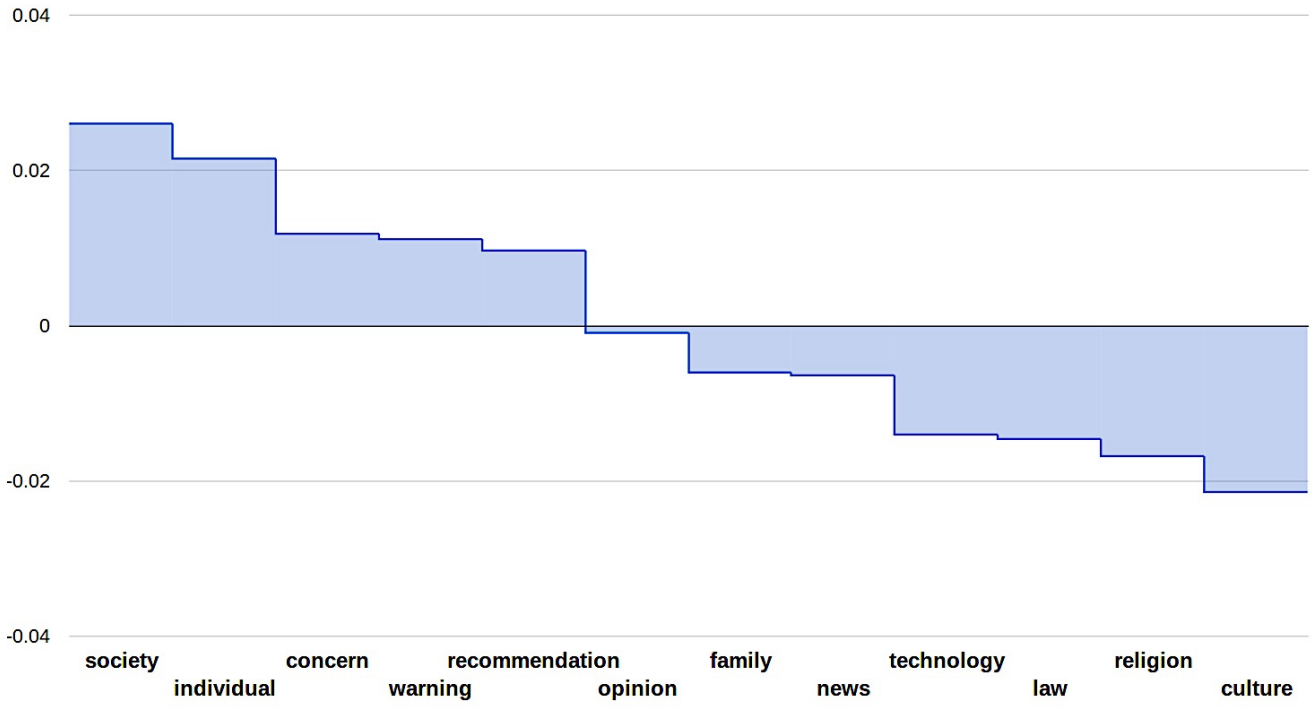

**Figure 5: Primary codes ranked by the difference between the fractions of women to men**

For each of the primary codes we computed the number of tweets posted by men and women, and then normalized by the total number of tweets posted by each gender to get the fractions of tweets contributed by men and women for each code. The difference between the fractions of women to men was then computed. Figure 5 shows the resulting fractions. Topics mostly discussed by women are to the left and to the right are topics mostly discussed by men.

The relatively low numbers on the vertical axis reveal that male and female participation is similarly distributed over the different codes. However, small—yet important—nuances are present. Figure 5 shows that women are more active on topics related to *society* and *individuals* while men are more active on the religious and cultural aspects of privacy. We also observe that women tend to express more *concerns* and *warnings* about their privacy than men. This is probably due to the weight of societal expectations on women to adhere to a specific code of conduct. This also shows that women in the GCC are aware of the role society assigned them as bearers of cultural norms; they are responsible for family honor. On the flip side, our results show that men had a higher tendency to participate in discussions of technical and legal nature. Lastly, the code *family* appears to rank higher amongst men, which can be explained in terms of their role in society as the bread winners and the protectors of the family. This could be seen in the often employed paternalistic language men use when discussing women's privacy.In sum, this discourse shows that Qatari men and women are both playing an active role when it comes to privacy, with different levels of interest regarding specific contexts.

## CONCLUSION

The results of this study expand and enrich the Western-centric privacy notion inscribed in system design by providing the perspective of privacy as practiced in the everyday life in Qatar. We contribute the following observations specific to the Qatari society (largely applicable to the rest of the GCC): (1) The notion and expression of privacy is highly gendered, such that the specific requirements for privacy differ greatly between males and females. Surprisingly, in our dataset, women were as involved in the definition of these requirements as men. (2) Privacy includes what people are exposed to, not only what they are themselves exposing. That is, one's environment—the behavior of others around them—reflects on one's own honor, making privacy a highly communal concept. (3) An individual's privacy should be preserved even after death, in accordance with the Muslim belief in the afterlife. (4) Privacy is a moral value highly influenced by religion. It is defined in religious texts and by the cultural norms that have been largely shaped by Islam.

This research offers new insights into the ways Qataris conceptualize and enact privacy and honor. We also present situations in which privacy and honor are challenged or compromised. The discourse of privacy and honor on Twitter reveals the unique perspectives and methods Qataris use to mindfully negotiate these values and communicate them to others. Thus, this method captures daily definitions of privacy in the online community, some of which may be impossible to accurately capture through traditional survey methods. That said, we believe that there is an opportunity to enrich this study by including additional qualitative data.

Privacy and honor are closely connected in the context of the GCC. Our three comprehensive mixed-method analyses of Twitter communications further highlight the multifaceted nature of these ever-evolving—but traditionally entrenched—values that have multiple life spans. Future work will involve more contextually grounded understandings of values and value enactment, and how these actions impact technology use and design. In line with [16,28,35,38], we call for more research on cross-cultural value tensions, as this is crucial for the development of privacy-aware systems that speak to a global audience.